\documentclass[epj,final]{svjour}

\usepackage{amsmath}
\usepackage{bm}
\usepackage{graphicx}
\usepackage[bookmarks=false,pdfstartview=FitH]{hyperref}

\renewcommand{\Im}{\operatorname{Im}}
\newcommand{\sgn}{\operatorname{sgn}}
\newcommand{\abs}[1]{\lvert#1\rvert}
\newcommand{\pd}[2]{\frac{\partial#1}{\partial#2}}
\newcommand{\bsigma}{\bm\sigma}

\begin{document}

\title{Adiabatic quantum pumping in graphene with magnetic barriers}

\author{E. Grichuk \and E. Manykin}
\institute{National Research Center ``Kurchatov Institute'', Kurchatov Sq.~1, Moscow 123182, Russia}

\abstract{We study an adiabatic quantum pump effect in a two terminal graphene device with two oscillating square electric barriers and a stationary magnetic barrier using the scattering matrix approach. The model employs the low-energy Dirac approximation and incorporates the possible existence of a finite band gap in graphene spectrum. We show that in this case valley-polarized and pure valley currents can be pumped due to the valley symmetry breaking. For a $\delta$-function magnetic barrier we present analytical expressions for bilinear total and valley pumping responses. These results are compared to numerical ones for a double $\delta$-function, a square and a triple square magnetic barriers.}

\maketitle

\section{Introduction}

Graphene, a single atomic layer of carbon atoms packed into a honeycomb lattice, is now being intensively studied both theoretically and experimentally. In an endeavor to construct new nanoelectronic devices based on graphene, a particular attention of researchers is paid to its unconventional and somewhat counterintuitive electronic properties. Graphene electronic spectrum comprises two valleys, referred to as $K$ and $K'$, and in the low energy approximation electron dynamics in each valley is governed by a 2D Dirac equation for massless particles. Thus, quasiparticles in graphene have linear spectrum with the Fermi velocity $v_F \approx c/300$. This results in a peculiar behavior of electrons in this material, which manifest itself in the Klein tunnelling~\cite{ANS98,KNG06}, the half-integer quantum Hall effect~\cite{NGM+05}, the Veselago lensing of electrons~\cite{CFA07} and many other effects~\cite{CNGP+09,Per10,AAB+10,DAHR11}.

When graphene is considered as a basis for nanoelectronics (e.\,g., graphene FETs), the Klein tunneling through a potential barrier is of special interest. Touching of conduction and valence bands and the existence of propagating modes below the barrier allow electrons to penetrate it with high probability (with unit probability at normal incidence). Thus, this effect hampers the confinement of electrons in graphene-based field effect devices. A number of ways have been proposed to circumvent this obstacle. For example, a graphene sheet can be cut to form nanoribbons, nanoislands, nanorings and other nanostructures. Operating principles of many suggested graphene devices based on such structures crucially depend on the atomically precise edge configuration, e.\,g., a zigzag or an armchair edge of graphene nanoribbons. Their fabrication is an experimentally challenging task and forces one to look for alternative methods.

Let us briefly mention two of them which are relevant for the present paper. First, electron confinement in graphene can be achieved by means of inhomogeneous magnetic field~\cite{DMDE07,RVMP08,RVP11} that can be generated, in particular, by depositing ferromagnetic strips (gates) on the top of a dielectric layer covering a graphene flake. Similar technique is now widely utilized to create magnetic barriers in structures with conventional two-dimensional electron gas. Their graphene counterparts are expected to be realizable in the near future with various useful applications, for example, in graphene-based spintronics. Second, a special choice of a substrate can break graphene sublattice symmetry by introducing staggered sublattice potential and induce a finite energy band gap~\cite{ZGF+07,GKB+07}. Depending on a substrate (e.\,g., $h$-BN or SiC), values from a few tens of meV up to a few tenths of eV have been reported. The combination of an external magnetic field and an induced band gap can be used to generate and detect valley-polarized currents in graphene---a necessary step toward the so-called graphene ``valleytronics'' (valley-based electronics), which exploits peculiar distributions of carriers in the valley space~\cite{XYN07,ZC12,Zha12,SHH12}.

Most research of electronic transport in graphene is focused on stationary problems. A variety of new effects emerges when one considers non-stationary ones. An interesting phenomenon, initially due to D.\,J.\,Thouless~\cite{Tho83}, is a quantum pump effect, in which a periodic modulation of parameters of a quantum system produces a finite dc~current through it even in the absence of an external bias. Quantum pumping in graphene has recently attracted increasing attention of researchers~\cite{ZC09,PSS09,ZL11,GM10,WC10,TB10,GM11,GM11b,ZCL11,ZLLC12,ZLC12,AB11,SPKS11,WCC12}. The unusual electronic spectrum of graphene was demonstrated to have a significant impact on the effect. In particular, the important role of the Klein tunneling and evanescent modes was stressed~\cite{ZC09,PSS09,ZL11}. The potential use of a quantum pump effect in graphene-based spintronics was also discussed~\cite{WC10,TB10,GM11,GM11b,ZCL11,ZLLC12,ZLC12}.

The interaction of graphene with laser field falls under the same category of time-dependent phenomena, and has been intensively studied~\cite{CPRT11,AC09,SA11,GFAA11,KOB+11}. It was demonstrated that external radiation field can qualitatively change the electronic transport properties of graphene. In particular, it can be employed for tuning a band gap in graphene~\cite{CPRT11}, for generating valley-polarized current in bilayer graphene~\cite{AC09} and for analysing transport in graphene superlattices using a spatial-temporal duality between static spatially periodic electric and magnetic fields and time-periodic laser field~\cite{SA11}. It is also interesting to note that laser field can induce a topologically nontrivial band gap in (otherwise gapless) graphene spectrum leading to the formation of a topological insulator state in graphene~\cite{KOB+11}.

Motivated by possible applications of a quantum pump effect in graphene valleytronics, in this paper, we extend previous studies of quantum pumping in graphene by taking into consideration the valley degree of freedom of electrons.

\section{Delta-function magnetic barrier}

\begin{figure}
	\includegraphics[width = .99\columnwidth]{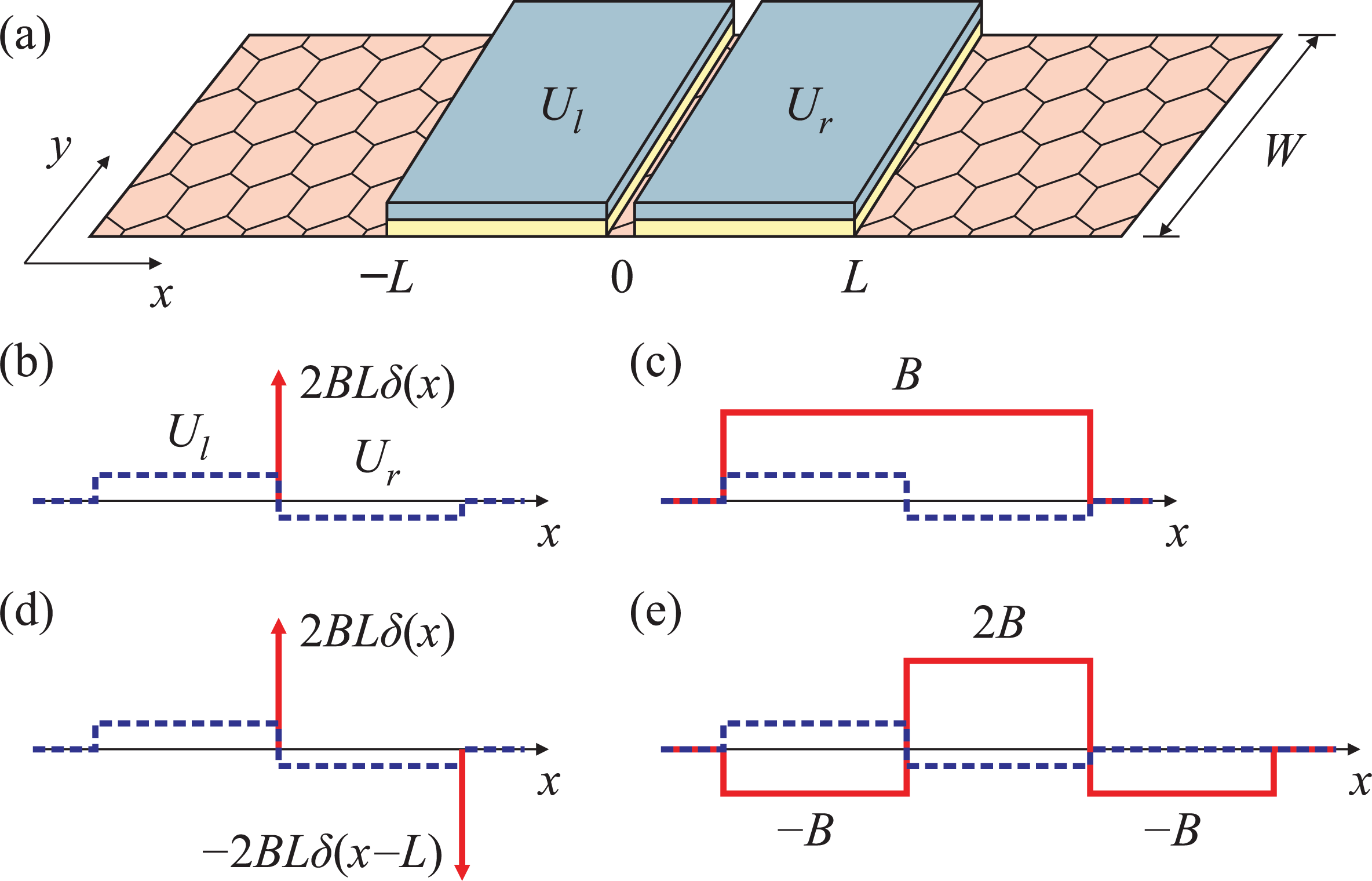}
	\caption{(a)~A schematic structure of a proposed graphene device. The device is formed by a wide graphene ribbon with two square electric barriers (produced by top metallic gates) of heights~$U_l$ and $U_r$ and width~$L$ as well as (b)~a single $\delta$-function magnetic barrier or (c)~a square magnetic barrier of width~$2L$ or (d)~a double $\delta$-function magnetic barrier or (e)~a triple square barrier of width~$3L$.}
	\label{fig1}
\end{figure}

The system that we examine is a standard two terminal quantum pump device that is formed by a wide graphene strip of width~$W$ in the $(x,y)$-plane with two electric barriers whose heights~$U_l$ and~$U_r$ can be periodically modulated in time and one stationary magnetic barrier (Fig.~\ref{fig1}). Experimentally such barriers can be realized by depositing ferromagnetic strips on top of a graphene ribbon~\cite{MPV94}. In this paper we limit ourselves to considering simplest profiles of a magnetic barrier: $\delta$-function barriers and square barriers.

To model the electron transport in the graphene device, we employ the low-energy Dirac approximation. The electronic transport is assumed to be completely phase-coherent between the leads. The single-valley Hamiltonian of the device reads
\begin{equation}
\label{eq.HamiltonianNoGap}
H = v_F \bsigma [\vec{p} + \vec{A}(x)] + U(x),
\end{equation}
where $v_F$ is the Fermi velocity in graphene, $\vec{p}$ is the canonical momentum operator, $\vec{A}(x)$ is the vector potential corresponding to a magnetic barrier and $U(x)$ is the scalar potential corresponding to electric barriers. To simplify the notation, we put $\hbar = e = 1$.

For electric barriers we adopt the following profile, which is translationally invariant in the $y$-direction:
\begin{equation}
U(x) = \begin{cases} U_l, & -L \leq x < 0, \\ U_r, & 0 \leq x \leq L, \\ 0, & \text{otherwise}. \end{cases}
\end{equation}

The sharp edge approximation, which is suitable for analytical calculations, to physically relevant smooth barriers is justified if the edge smearing length of barriers is smaller than the Fermi wavelength of electrons, being at the same time larger than the graphene lattice constant to suppress the intervalley scattering.

To obtain an analytical solution, we first consider a single $\delta$-function magnetic barrier $\vec{B}(x) = 2BL\delta(x)\hat{\vec{z}}$, translationally invariant in the $y$-direction with magnetic field perpendicular to the graphene sheet and localized between electric barriers (Fig.~\ref{fig1}b). In the Landau gauge the corresponding vector potential takes the form $\vec{A}(x) = A(x)\hat{\vec{y}}$ with
\begin{equation}
\label{eq.DeltaMagnBarrier}
A(x) = \begin{cases} -BL, & x < 0, \\ BL, & x > 0. \end{cases}
\end{equation}

For later use, we define the following length and energy scales inherent to the problem: $\ell_B = 1/\sqrt{B}$ and $E_L = v_F/L$. For a typical magnetic field $B_0 = 0.1$~T and Fermi velocity $v_F = 0.54$~eV$\cdot$nm, we have $L_0\equiv\ell_{B_0} = 81.1$~nm and $E_{L_0} = 6.6$~meV. The Zeeman splitting $E_Z = g\mu_B B_0 = 1.8 \cdot 10^{-3} E_{L_0}$ is small and will be neglected here (the spin degeneracy factor of~2 is omitted throughout the paper).

The eigenfunctions of the Hamiltonian~\eqref{eq.HamiltonianNoGap} are well known~\cite{CNGP+09,Per10}. Due to the translational invariance of the system in the $y$-direction, in the region~$\alpha$ where $U(x) = U_\alpha$ and $A(x) = A_\alpha$, they can be written as $\psi_\alpha(x,y) = e^{iqy} \psi_\alpha(x)$ with
\begin{equation}
\label{eq.EigenFunctionsNoGap}
\psi_\alpha(x) = \left\{\frac{1}{\sqrt{k_\alpha}} \begin{pmatrix} 1 \\ \eta \end{pmatrix} e^{ik_\alpha x},\ \frac{1}{\sqrt{k_\alpha}} \begin{pmatrix} 1 \\ -\eta^* \end{pmatrix} e^{-ik_\alpha x}\right\},
\end{equation}
where $\eta = v_F(k_\alpha + i q_\alpha)/(E - U_\alpha)$. The eigenenergy~$E$ is given by $E = U_\alpha \pm v_F \sqrt{k_\alpha^2 + q_\alpha^2}$. For electron-like states ($E > U_\alpha$) the first and the second states in~\eqref{eq.EigenFunctionsNoGap} correspond to left- and right-moving carriers, respectively. Below only electron-like states in the leads ($E > 0$) are considered. Note that for finite magnetic field, $k_l \ne k_r$, and the factor~$1/\sqrt{k_\alpha}$ is included to ensure the unitarity of the scattering matrix. The transverse canonical momentum~$q$, which is conserved in the scattering process, and kinetic momentum~$q_\alpha$ are related by $q_\alpha = q + A_\alpha$. In the left ($x < 0$) and in the right ($x > 0$) regions we have, respectively,
\begin{equation}
q_l = q - BL \quad\text{and}\quad q_r = q + BL.
\end{equation}

Transverse modes indexed by~$q$ are not mixed by the scattering, and the scattering problem is solved independently for each mode~$q$. This can be done conveniently using a transfer matrix method~\cite{NN09}. In this method potential profiles are approximated by piecewise constant functions. A transfer matrix of the whole system is then calculated as a product of (known) transfer matrices describing the propagation of an electron through regions with uniform potentials and potential steps. A scattering matrix~$S$ can be extracted from a transfer one using well-known relations.

The device considered in our study is up-down symmetric, i.\,e.\ it possesses a symmetry axis parallel to the current direction (the $x$-direction). Then, the $S$-matrix is always symmetric~\cite{BM96} and can be cast in the form
\begin{equation}
\label{eq.Smatrix}
S = \begin{pmatrix} r & t' \\ t & r' \end{pmatrix} = e^{i\gamma} \begin{pmatrix} \sqrt{1-T}e^{i\varphi} & i\sqrt{T} \\ i\sqrt{T} & \sqrt{1-T}e^{-i\varphi} \end{pmatrix},
\end{equation}
where $r$ and $t$ ($r'$ and $t'=t$) are reflection and transmission amplitudes for electrons incident from the left (right) lead.

Suppose now that electric barrier heights~$U_l$ and $U_r$ are periodically varied in time so that a point $(U_l(t), U_r(t))$ traverses a closed contour~$C_U$ in the $(U_l, U_r)$ plane, say,
\begin{equation}
\begin{split}
U_l(t) &= U_{l0} + \delta U_l \sin(\omega t), \\
U_r(t) &= U_{r0} + \delta U_r \sin(\omega t - \phi),
\end{split}
\end{equation}
with some fixed phase shift~$\phi$. This will lead to a finite charge being pumped through a device per each cycle. If the pumping frequency is small enough (adiabatic regime), the transported charge in the mode~$q$ can be expressed via a ``frozen'' scattering matrix $S(t) = S(U_l(t), U_r(t))$, which depends on the time~$t$ as a parameter~\cite{Mos11,Bro98}:
\begin{equation}
\label{eq.Q}
Q(E, q) = \int_{C_U}du_l du_r \, \Pi(E, q, U_l, U_r), \quad u_\alpha = \frac{U_\alpha}{E_L},
\end{equation}
where
\begin{multline}
\label{eq.Pi}
\Pi(E, q, U_l, U_r) = \frac{E_L^2}{\pi} \Im \left( \frac{\partial r^*}{\partial U_l} \frac{\partial r}{\partial U_r} + \frac{\partial t^*}{\partial U_l} \frac{\partial t}{\partial U_r} \right) = \\
= \frac{E_L^2}{2\pi} \left( \pd{\varphi}{U_l} \pd{T}{U_r} - \pd{T}{U_l}\pd{\varphi}{U_r} \right).
\end{multline}

The expression~\eqref{eq.Q} can be significantly simplified provided the amplitudes~$\delta U_\alpha$ are small ($\delta U_\alpha \ll E_L$) so that $\Pi(E, q, U_l, U_r)$ is approximately constant within~$C_U$. Then, the response of a quantum pump is bilinear in the amplitudes~$\delta U_l$ and $\delta U_r$:
\begin{equation}
Q(E, q) = \Pi(E, q, U_{l0}, U_{r0}) A_u, \quad A_u = \frac{A_U}{E_L^2},
\end{equation}
where $A_U = \pi \delta U_l \delta U_r \sin\phi$ is the area enclosed by the contour~$C_U$ on the $(U_l, U_r)$~plane.

To characterize the pumping, we numerically calculate the total pumped charge~$Q(E)$ and the pumping response~$\chi(E)$ averaged over the transverse modes
\begin{equation}
Q(E) = \frac{A_U}{E_L^2} \sum_q \Pi(E, q), \ \chi(E) = \frac{1}{N_m} \sum_q \Pi(E, q),
\end{equation}
where summation extends over $N_m$~transverse modes that are propagating in both leads at the energy~$E$. For wide ribbons the sums can be approximated by integrals.

A succinct analytical solution is obtained for a device with $U_{l0} = U_{r0} = 0$. For the transmission $T = \abs{t}^2$ through the $\delta$-function magnetic barrier~\eqref{eq.DeltaMagnBarrier} in the absence of electric barriers one gets~\cite{RVMP08}
\begin{equation}
\label{eq.Transmission}
T(E, q) = \frac{4k_l k_r}{(k_l+k_r)^2+(2L/\ell_B^2)^2},
\end{equation}
where $k_\alpha = \sqrt{E^2/v_F^2 - q_\alpha^2}$, $\alpha = l,r$. The magnetic barrier has a finite transparency only if the energy~$E$ exceeds a minimum value~\cite{DMDE07}
\begin{equation}
\label{eq.Emin}
E_\mathrm{m} = \left(\frac{L}{\ell_B}\right)^2 E_L,
\end{equation}
otherwise propagating modes do not exist in both leads.

Taking the derivatives of the scattering matrix~$S$ in \eqref{eq.Pi}, we end up with the following expression for the pumping response:
\begin{multline}
\label{eq.PiNoGap}
\Pi(E, q) = \frac{2}{\pi} \frac{T(q_l L)(q_r L)}{(k_l L)^2(k_r L)^2} \sin(k_l L)\sin(k_r L) \times \\
\times \sin[(k_l + k_r) L] + \frac{2}{\pi} \left(\frac{LE}{\ell_BE_L}\right)^2 \times \\
\times \frac{T^2\sgn(B)}{(k_l L)^2(k_r L)^2} \left[ \frac{q_r}{k_r} \sin^2(k_r L) - \frac{q_l}{k_l} \sin^2(k_l L) \right],
\end{multline}
where $\sgn(z)$ is a sign function and a factor of~2 was included to account for the double valley degeneracy (see next section).

The second term in this expression contains the factor $1/\ell_B^2 = B$ and gives no contribution in the limit of vanishing magnetic field. The expression then simplifies to the result obtained by E.\,Prada et~al.~\cite{PSS09}
\begin{equation}
\label{eq.PiNoMF}
\Pi(E, q) = \frac{2}{\pi} \frac{2(qL)^2 \sin^3(kL)\cos(kL)}{(kL)^4}
\end{equation}
with $k = k_l = k_r$ and $q = q_l = q_r$. It is readily seen that the response vanishes for the modes with~$q=0$, which travel normally to the barriers. These modes are insensitive to the electric barriers~$U_l$ and~$U_r$ due to the Klein paradox and therefore cannot be pumped~\cite{PSS09}. This can also be seen from~\eqref{eq.Pi}. In the absence of magnetic field $\partial T/\partial U_l = \partial T/\partial U_r$, and the pumped current vanishes if the transmission~$T$ is independent of~$U_l$ and $U_r$.

The Klein tunneling also manifests itself for a magnetic barrier of finite height. The first term in~\eqref{eq.PiNoGap} contains the product~$q_lq_r$ of the transverse kinetic momenta. Hence, this term drops out if the mode propagates in the left or in the right lead normally to the barrier. In either case, the transmission is insensitive to the height of the corresponding barrier. The contribution of the second term is always finite and indicates the phase coherent nature of a quantum pump effect: although one of the barriers does not affect the transmission probability of an electron, it influences its phase~\cite{Mos11}. If, say, $q_l = 0$, then $\partial T/\partial U_l = 0$, but $\partial T/\partial U_r$ remains finite, and for the pumping response we have
\begin{equation}
\Pi(E, q) = -\frac{E_L}{\pi} \pd{T}{U_r},
\end{equation}
where a simple relation $\partial\varphi/\partial U_l = -L/v_F = -1/E_L$ for the mode with $q_l = 0$ was used.

The spatial symmetry of a device can impose certain symmetries on the response function~$\Pi(E, q)$. It can be easily demonstrated that the Hamiltonian~\eqref{eq.HamiltonianNoGap} satisfies the symplectic symmetry $\mathcal{S} H(B) \mathcal{S}^{-1} = H(-B)$ (operation of time-reversal in a single valley), where $\mathcal{S} = i\sigma_y \mathcal{C}$ with $\mathcal{C}$ the operator of complex conjugation. This symmetry implies that~\cite{AAK00,SAA00,WK03}
\begin{equation}
\Pi(q, B) = \Pi(-q, -B)
\end{equation}
and, therefore,
\begin{equation}
Q(B) = Q(-B), \quad \chi(B) = \chi(-B).
\end{equation}
These relations are valid for arbitrary $U_{l0}$ and $U_{r0}$. If the electric barrier is left-right symmetric ($U_{l0} = U_{r0}$) and the vector potential is left-right antisymmetric ($A(x)=-A(-x)$), then the pumping response is also invariant with respect to the inversion of~$q$ alone:
\begin{equation}
\Pi(q, B) = \Pi(-q, B).
\end{equation}
This equation is a consequence of the symmetry $\mathcal{O} H(B) \mathcal{O}^{-1} = H(B, U_r \leftrightarrow U_l)$, where $\mathcal{O} = \sigma_z R_x R_y$ with $R_x$ ($R_y$) the reflection about the $x$-axis, $y \to -y$ ($y$-axis, $x \to -x$) and $U_r \leftrightarrow U_l$ denotes the interchange of the electric barriers heights.

\begin{figure}
	\includegraphics[width = .99\columnwidth]{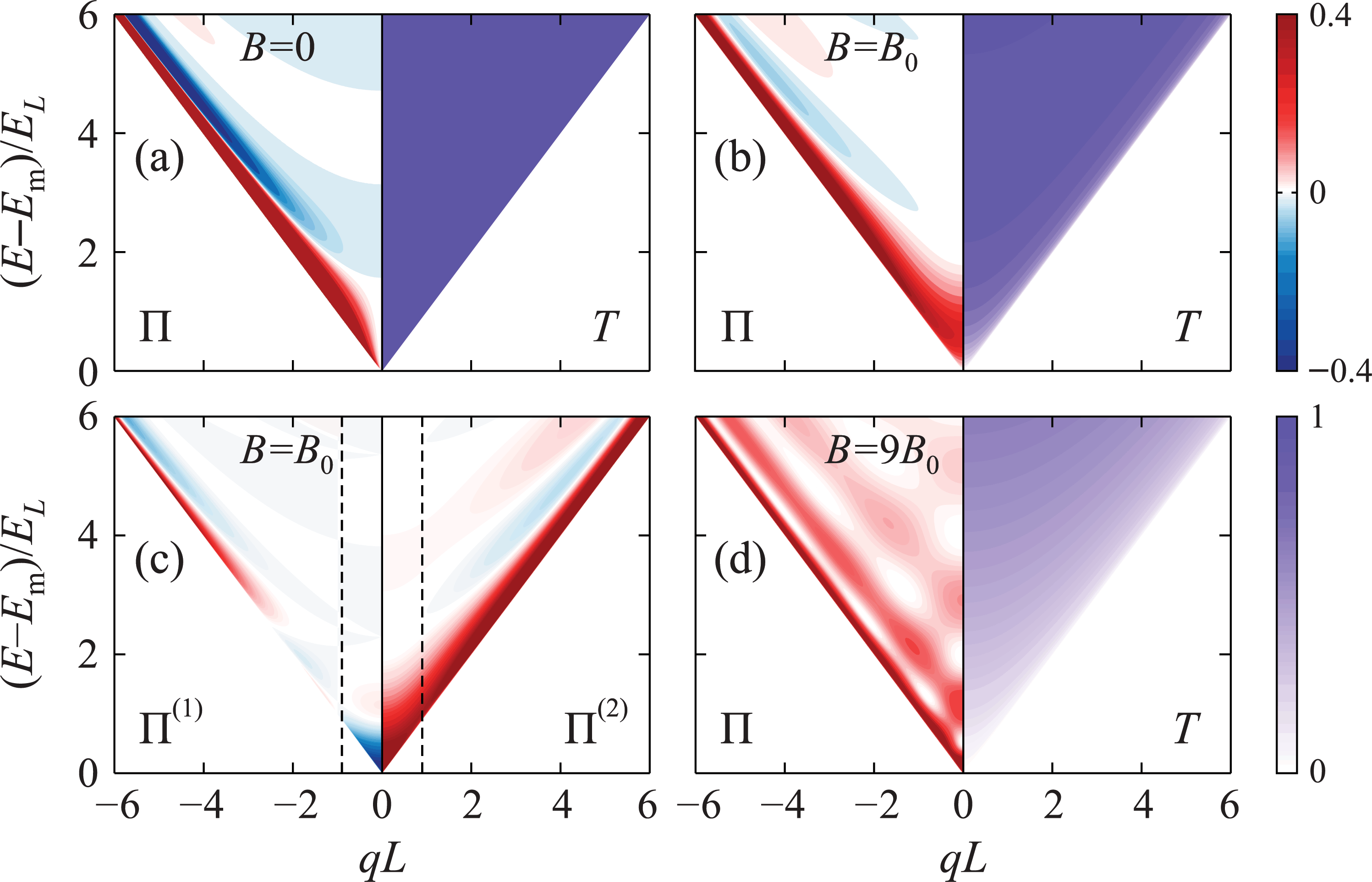}
	\caption{(a)--(d) Contour plot of the pumping response~$\Pi(E,q)$ and the transmission~$T(E,q)$ for the $\delta$-function magnetic barrier~\eqref{eq.DeltaMagnBarrier} in graphene with gapless spectrum as a function of the canonical momentum~$q$ and the Fermi energy~$E$. In panel~(c) the contributions of the first ($\Pi^{(1)}$) and the second ($\Pi^{(2)}$) terms in~\eqref{eq.PiNoGap} are shown separately. Dashed vertical lines correspond to $q_l = 0$ and $q_r=0$. All distributions are even functions of~$q$, and only half of each distribution is shown. Width of electric barriers is $L = L_0$.}
	\label{fig2}
\end{figure}

The dependences of the pumping response~$\Pi(E, q)$ and the transmission~$T(E, q)$ on the canonical momentum~$q$ and the Fermi energy~$E$ are shown in Fig.~\ref{fig2}. One can observe that the perfect Klein tunneling in the vicinity of $q=0$ is destroyed by the external magnetic field. As it increases, an interference pattern becomes visible (Fig.~\ref{fig2}d). A similar pattern emerges in the transmission through the device with finite~$U_{l0}$ and $U_{r0}$ (not shown).

\begin{figure}
	\includegraphics[width = .99\columnwidth]{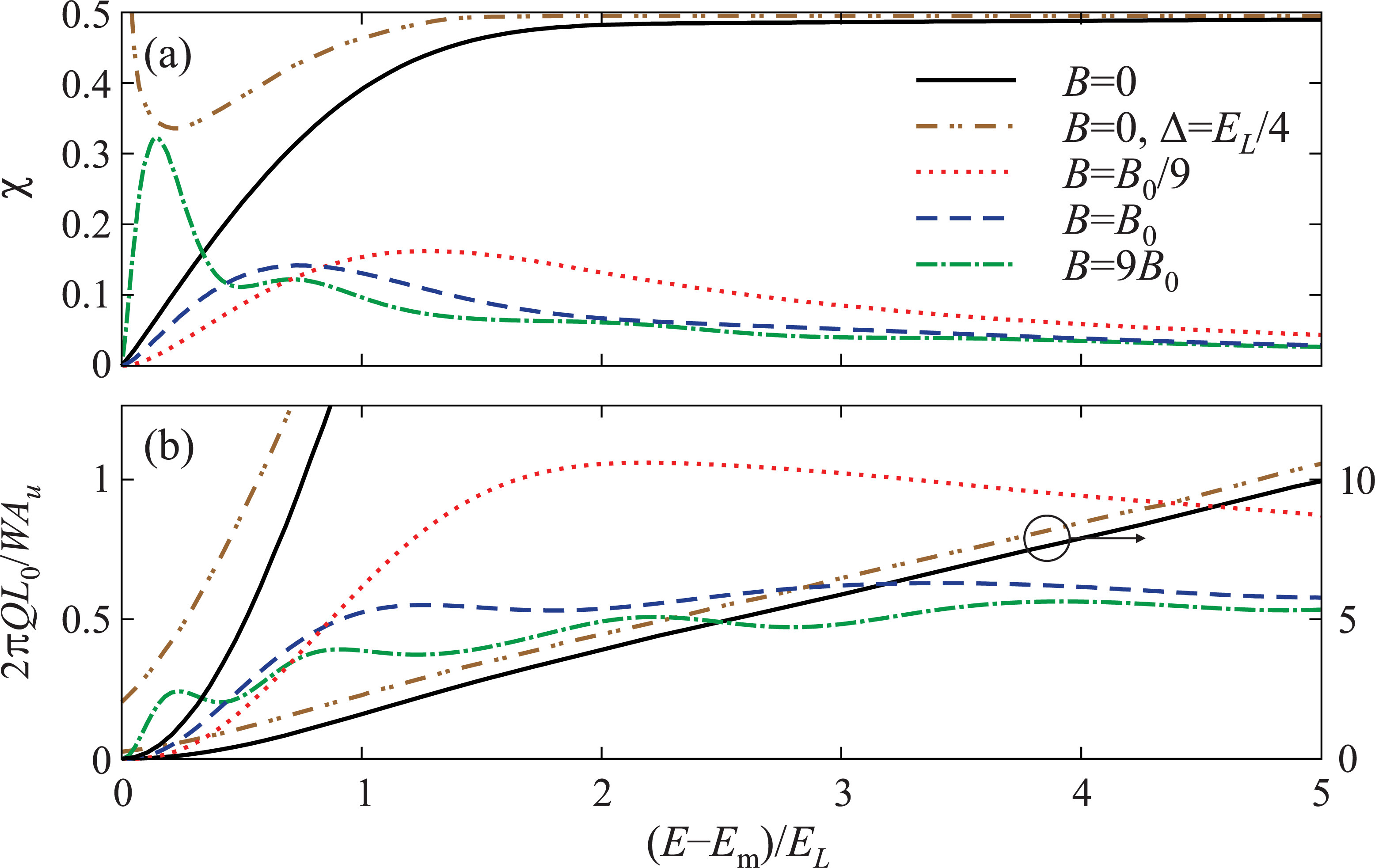}
	\caption{(a) The pumping response~$\chi(E)$ averaged over transverse modes and (b)~the pumped charge~$Q(E)$ for a wide ribbon as a function of the Fermi energy~$E$ for different magnetic field strengths.}
	\label{fig3}
\end{figure}

In Fig.~\ref{fig3} we plot the averaged pumping response~$\chi(E)$ and the total pumped charge~$Q(E)$ for different values of the magnetic field. In zero magnetic field the averaged response saturates at the value of~$0.5$~\cite{PSS09}. The magnetic field destroys the monotonic behavior and diminishes the effect at high Fermi energy. When~$B$ is swept from zero to $B_0$, the response is significantly reduced. At stronger magnetic field a similar but weaker tendency is observed.

In all curves in Fig.~\ref{fig3} the pumped charge~$Q(E)$ is a positive function of energy, i.\,e. the pumping is directed. However, by tuning the electric barrier heights~$U_{l0}$ and~$U_{r0}$ a regime, in which $Q(E)$ becomes a sign-changing function, can be achieved. Then, the pumping direction can be reversed by varying the Fermi energy---a generic feature inherent to quantum pumps. It can be used to generate pure spin~\cite{GM11,GM11b,ZCL11,ZLLC12,ZLC12} and pure valley (see below) currents.

\section{Valley-polarized current}

In the above analysis a valley degree of freedom was of no importance and was accounted of by a factor of~2. The symmetry between the $K$ and $K'$ valleys is destroyed in gapped graphene with broken inversion symmetry~\cite{XYN07,ZC12,Zha12}, and this opens the possibility to pump valley-polarized currents.

The graphene Hamiltonian that describes the states in both valleys can be written in various unitarily equivalent representations. Choosing a basis $(\psi_A, \psi_B)^T$ in the $K$~valley and $(-\psi'_B, \psi'_A)^T$ the $K'$~valley the Hamiltonian takes the so-called valley isotropic form
\begin{equation}
\label{eq.HamiltonianGap}
H = v_F \bsigma [\vec{k} + \vec{A}(x) ] + \Delta \tau_z \sigma_z + U(x)
\end{equation}
with $\Delta$ the site energy difference between two sublattices (determined by the substrate) and $\tau_z=\tau$ the valley index: $\tau = 1$ $(-1)$ for the $K$ ($K'$) valley.

To obtain an analytical solution we again start with a $\delta$-function magnetic barrier~\eqref{eq.DeltaMagnBarrier}. Eigenfunctions of Hamiltonian~\eqref{eq.HamiltonianGap} with uniform scalar and vector potentials are given by~\eqref{eq.EigenFunctionsNoGap} with $\eta = v_F(k_\alpha + i q_\alpha)/(E - U_\alpha + \Delta\tau)$. From an eigenenergy $E = U_\alpha \pm \sqrt{\Delta^2 + v_F^2(k_\alpha^2 + q_\alpha^2)}$ it is seen that the spectrum has an energy gap~$2\Delta$. The minimum energy is now $E_\mathrm{m} = E_L\sqrt{(\Delta/E_L)^2+(L/\ell_B)^4}$. Using transfer matrix method and performing calculations similar to those in the previous section, we can find the transmission~$T_\tau(E, q)$ and the pumping response~$\Pi_\tau(E, q)$.

The expression~\eqref{eq.Transmission} for the transmission remains valid with $k_l$ and $k_r$ given by $k_\alpha = \sqrt{(E^2-\Delta^2)/v_F^2 - q_\alpha^2}$. The transmission is independent of the valley index~$\tau$, and no valley-polarized current is produced by applying a bias voltage. However, as was discussed in Refs.~\cite{ZC12,Zha12}, the transmission acquires a dependence on the valley index~$\tau$ in the presence of a non-uniform electric field. Hence, in contrast to the transmission, the pumping response for a device with $U_{l0} = U_{r0} = 0$ is expected to be valley-dependent. After tedious but straightforward calculations we arrive at
\begin{multline}
\label{eq.PiGap}
\Pi_\tau(E,q) = \frac{1}{\pi} \frac{T(q_lL)(q_rL)}{(k_lL)^2(k_rL)^2} \gamma_l \gamma_r \sin(k_lL)\sin(k_rL) \times \\
\times \sin[(k_l+k_r)L-\tau\phi] + \frac{1}{\pi} \left(\frac{LE}{\ell_BE_L}\right)^2 \sqrt{1-\left(\frac{\Delta}{E}\right)^2} \times \\
\times \frac{T^2\sgn(B)}{(k_lL)^2 (k_rL)^2} \left[ \frac{q_r}{k_r} \gamma_r \sin(k_rL) \sin(k_rL + \tau\phi_r) - \right. \\
\left. - \frac{q_l}{k_l} \gamma_l \sin(k_lL) \sin(k_lL - \tau\phi_l) \right],
\end{multline}
where
\begin{equation}
\sin\phi = \frac{E\Delta}{E_L^2} \frac{T\sgn(B)(k_l+k_r)L}{\gamma_l\gamma_r k_l k_r q_l q_r L^2 \ell_B^2}
\end{equation}
and
\begin{equation}
\gamma_\alpha = \sqrt{1 + \left(\frac{\Delta}{v_Fq_\alpha}\right)^2}, \quad 
\tan\phi_\alpha = \frac{k_\alpha\Delta}{q_\alpha E}.
\end{equation}

For vanishing band gap the expression reduces to the valley-independent result~\eqref{eq.PiNoGap} of the previous section. In the absence of a magnetic field we get the following generalization of~\eqref{eq.PiNoMF}:
\begin{equation}
\label{eq.PiGapNoMF}
\Pi(E, q) = \frac{2}{\pi} \frac{2\left[(qL)^2+(\Delta/E_L)^2\right]\sin^3(kL)\cos(kL)}{(kL)^4}.
\end{equation}

The presence of a band gap breaks the perfect Klein tunneling, and the contribution of modes with normal incidence (having $q=0$) to the pumping response becomes non-zero. Integrating~\eqref{eq.PiGapNoMF} at small~$k$, for the averaged response~$\chi(E)$ we obtain
\begin{equation}
\chi(E) = \frac{E^2 + \Delta^2}{2E_L\sqrt{E^2 - \Delta^2}}.
\end{equation}
Hence, in gapped graphene the averaged response~$\chi(E)$ diverges at small Fermi energy $E\to\Delta$ and the pumped charge $Q(E) \sim (E^2+\Delta^2)/E_L^2$ tends to a constant value (Fig.~\ref{fig3}).

To characterize the currents carried by electrons in different valleys, we introduce the total~$\Pi_c(E, q)$ and the valley~$\Pi_v(E, q)$ responses
\begin{equation}
\begin{split}
\Pi_c(E, q) &= \Pi_K(E, q) + \Pi_{K'}(E, q),\\
\Pi_v(E, q) &= \Pi_K(E, q) - \Pi_{K'}(E, q)
\end{split}
\end{equation}
and, analogously, the total~$Q_c(E)$ and the valley~$Q_v(E)$ pumped charges.

From \eqref{eq.PiGap} for the valley response we get
\begin{multline}
\Pi_v(E, q) = \frac{2}{\pi} \left(\frac{L}{\ell_B}\right)^2 \frac{E \Delta}{E_L^2} \frac{T^2\sgn(B)}{(k_lL)^2 (k_rL)^2} \times \\
\times \biggl\{-\frac{(k_l+k_r)L}{(k_lL)(k_rL)} \cos[(k_l+k_r)L] \sin(k_lL) \sin(k_rL) + \\
+ \sin[(k_l+k_r)L] \cos[(k_r-k_l)L] \biggr\}.
\end{multline}

This expression contains the factor~$B\Delta$, and as expected, no valley-polarized current is pumped in zero magnetic field or when the band gap closes.

From the above equations it is clear that the pumped charge~$Q_c$ is invariant with respect to the inversion of magnetic field, whereas the valley pumped charge~$Q_v$ changes sign:
\begin{equation}
\label{eq.QcQvSymm}
Q_c(B) = Q_c(-B), \quad Q_v(B) = -Q_v(-B).
\end{equation}

\begin{figure}
	\includegraphics[width = .99\columnwidth]{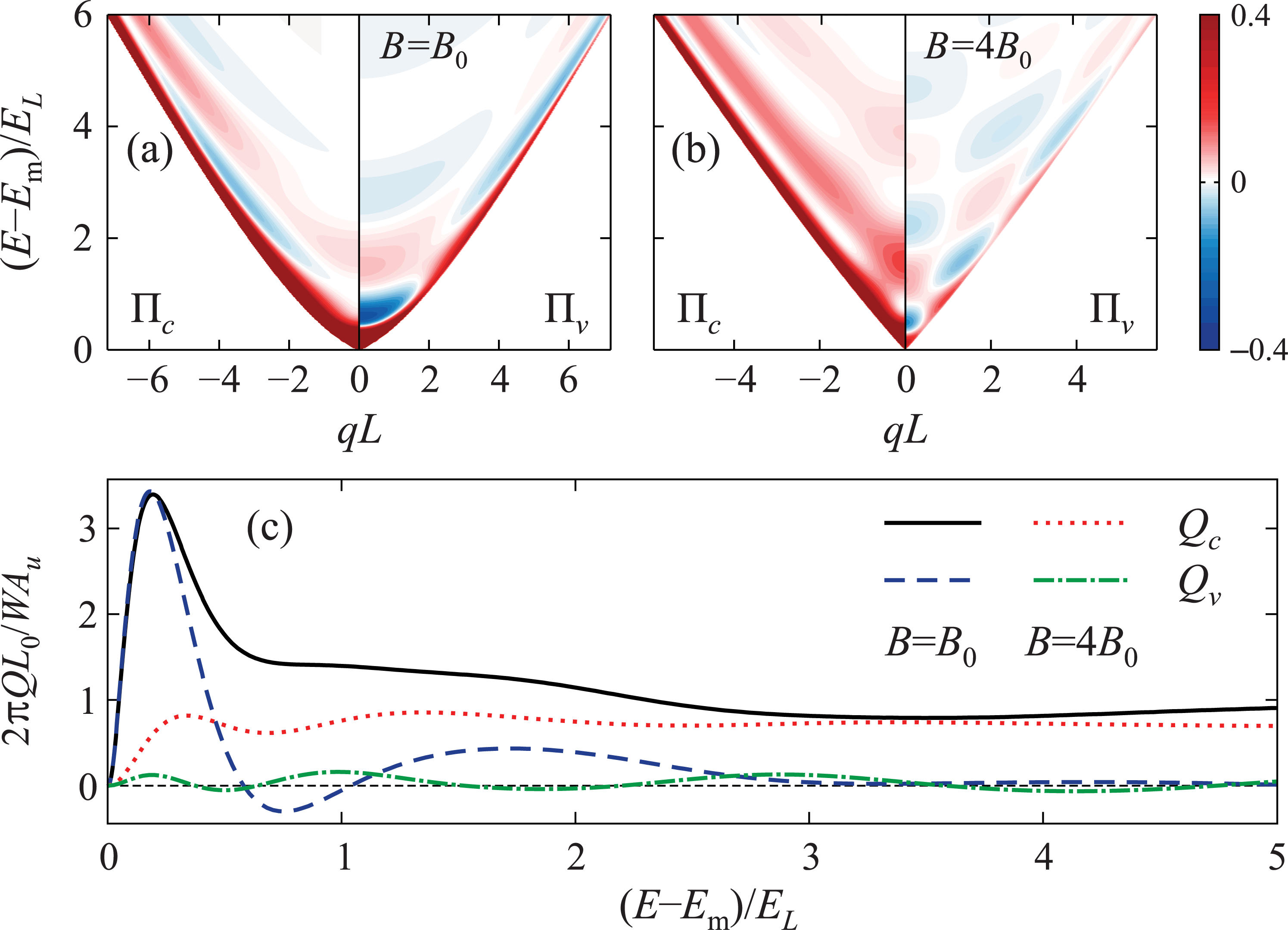}
	\caption{(a)--(b) Contour plot of the total~$\Pi_c(E,q)$ and the valley~$\Pi_v(E,q)$ pumping responses for the $\delta$-function magnetic barrier~\eqref{eq.DeltaMagnBarrier} in graphene with a gapped spectrum ($\Delta = 4E_L$) as a function of the canonical momentum~$q$ and the Fermi energy~$E$. (c)~The total~$Q_c(E)$ and the valley~$Q_v(E)$ pumped charges as a function of the Fermi energy~$E$ for different magnetic field strengths.}
	\label{fig4}
\end{figure}
These relations are dictated by the time reversal symmetry. The mass term~$\Delta\tau_z\sigma_z$ in the Hamiltonian~\eqref{eq.HamiltonianGap} breaks the symplectic symmetry~$\mathcal{S}$, but the orthogonal time-reversal symmetry $\mathcal{T} H(B) \mathcal{T}^{-1} = H(-B)$ with $\mathcal{T} = -\tau_y \sigma_y \mathcal{C}$ is preserved (here the Hamiltonian is viewed as a $4\times 4$~matrix acting in the valley and the sublattice spaces, and the operator~$\tau_y$ interchanges the $K$ and $K'$ valleys). It implies that
\begin{equation}
\Pi_\tau(q, B) = \Pi_{-\tau}(-q, -B),
\end{equation}
and the relations~\eqref{eq.QcQvSymm} follow immediately. The relation $\Pi_\tau(q, B) = \Pi_\tau(-q, B)$ that holds if the electric barrier is left-right symmetric ($U_{l0} = U_{r0}$) is not altered by the mass term and remains valid.

In Fig.~\ref{fig4} the pumping responses and the pumped charges are shown. The valley pumping response is appreciable at low Fermi energy and tends to zero at high energy. It is also gradually destroyed by higher magnetic field (note that the minimum Fermi energy $E_\mathrm{m} \sim B$ for high value of $B$). By changing the value of the magnetic field and the band gap a high polarization of the valley current can be obtained---a regime when current is mainly carried by electrons from one valley, so that $\abs{Q_c(E)} \approx \abs{Q_v(E)}$.

We now discuss a square magnetic barrier (with the same area) whose position coincides with that of the electric barriers (Fig.~\ref{fig1}c):
\begin{equation}
\label{eq.SquareMagnBarrier}
A(x) = \begin{cases} -BL, & x < -L, \\ Bx, & \abs{x} \leq L,\\ BL, & x > L. \end{cases}
\end{equation}

In order to apply the transfer-matrix method, we need eigenstates in the region with homogeneous electric field $U(x)=U_\alpha$ and magnetic field $B(x) = B$. The eigenstates can be expressed via parabolic cylinder functions $D_p(z)$ (sometimes also termed Weber functions)~\cite{GST06}:
\begin{equation}
\psi_\alpha(x) = \left(\begin{array}{c} D_{p-1}(\pm z) \\ \pm i\sqrt{2}\epsilon_\tau^{-1} D_p(\pm z) \end{array}\right),
\end{equation}
where
\begin{equation}
\begin{gathered}
z=\sqrt{2} \left(q\ell_B + \dfrac{x}{\ell_B}\right), \quad \epsilon_\tau = \frac{\ell_B}{L}\frac{E-U_\alpha+\Delta\tau}{E_L}, \\
p = \left(\frac{\ell_B}{L}\right)^2 \frac{(E - U_\alpha)^2 - \Delta^2}{2E_L^2}.
\end{gathered}
\end{equation}

To obtain the transmission~$T_\tau(E, q)$ and the pumping response~$\Pi_\tau(E, q)$ we resort to numerical evaluation of the transfer matrix. Similar to a $\delta$-function barrier, for a device without electric barriers the transmission turns out to be the same for both valleys due to the symmetry reasons. The valley dependence appears in the structure with non-vanishing electrostatic barriers~\cite{ZC12,Zha12}.

\begin{figure}
	\includegraphics[width = .99\columnwidth]{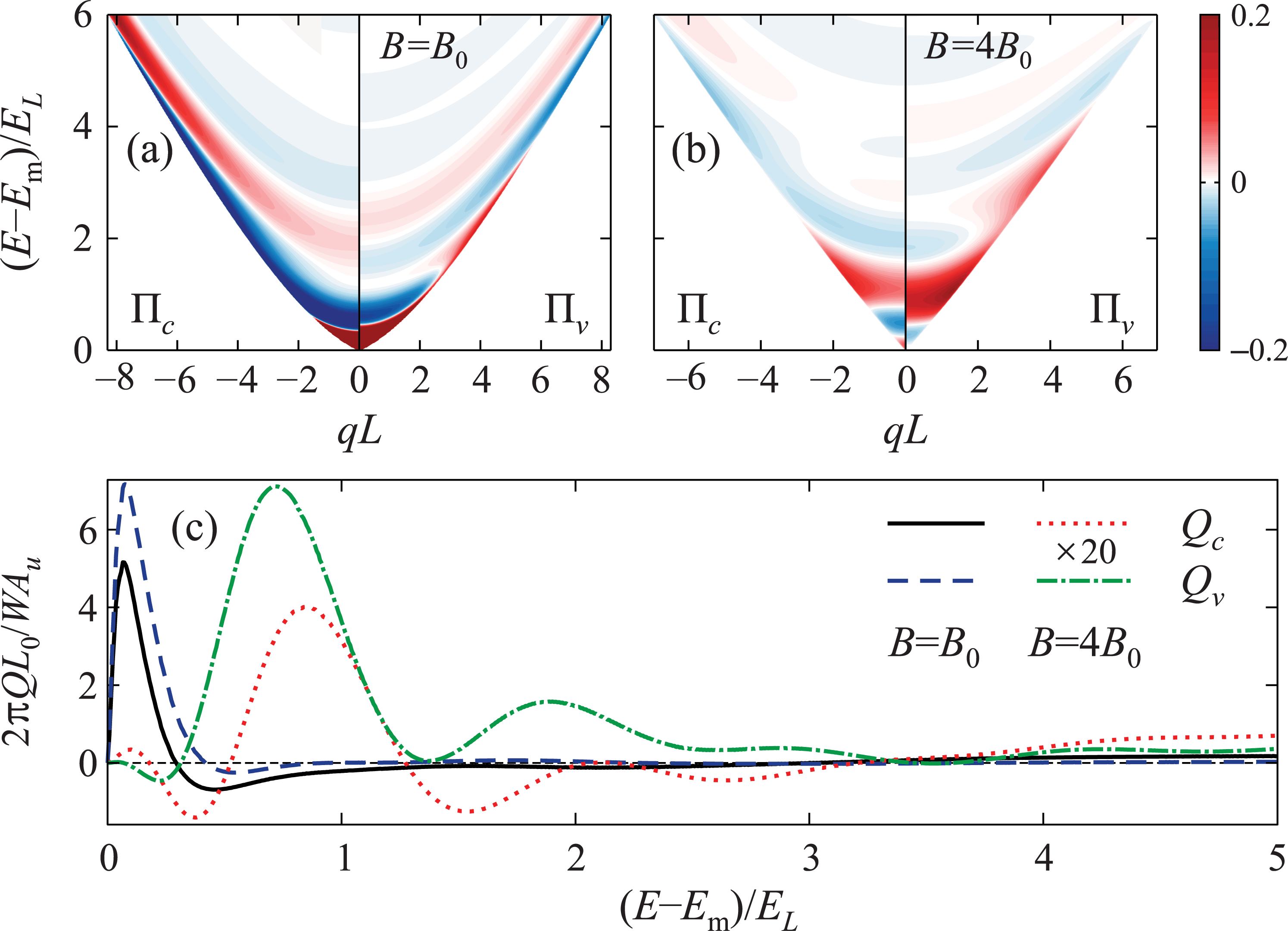}
	\caption{(a)--(b) Contour plot of the total~$\Pi_c(E,q)$ and the valley~$\Pi_v(E,q)$ pumping responses for the square magnetic barrier~\eqref{eq.SquareMagnBarrier} in graphene with a gapped spectrum ($\Delta = 4E_L$) as a function of the canonical momentum~$q$ and the Fermi energy~$E$. (c)~The total~$Q_c(E)$ and the valley~$Q_v(E)$ pumped charges as a function of the Fermi energy~$E$ for different magnetic field strengths. For clarity, the curves corresponding to $B=4B_0$ are multiplied by a factor of~20.}
	\label{fig5}
\end{figure}

For a finite width magnetic barrier the pumping response and the pumped charge show similar oscillatory behavior, as can be inferred from Fig.~\ref{fig5}. An interesting feature can be observed in Fig.~\ref{fig5}c. The total pumped charge~$Q_c(E)$ is a sign-changing function of the Fermi energy~$E$, so that at certain values of~$E$, the total pumped charge vanishes, whereas the valley charge remains finite. In this regime the currents carried by electrons from the valleys~$K$ and $K'$ flow in the opposite directions: $Q_K(E) = -Q_{K'}(E)$. They compensate each other (no net charge transport), and a pure valley current is generated. It is similar in spirit to a pure spin current generation in quantum pumps.

\section{Double delta-function and triple square magnetic barriers}

The magnetic field profiles created with ferromagnetic strips with magnetization perpendicular or parallel to the graphene plane are characterized with zero average magnetic field $\langle B_z \rangle = 0$~\cite{MPV94}. We consider two profiles having this property: a double $\delta$-function barrier with the magnetic field in the opposite directions and a triple square barrier.

Suppose the left gate is non-magnetic and the right one has magnetization parallel to the graphene plane. Then, the resulting magnetic field can be approximated with $\vec{B}(x) = 2BL\left[\delta(x) - \delta(x-L)\right] \hat{\vec{z}}$ (Fig.~\ref{fig1}d). The corresponding vector potential takes the form
\begin{equation}
\label{eq.DoubleDeltaMagnBarrier}
A(x) = \begin{cases}0, & x<0\ \text{or}\ x > L\\ 2BL, & 0 \leq x \leq L.\end{cases}
\end{equation}
To model the magnetic field produced by the right gate with perpendicular magnetization, we employ a triple square profile (Fig.~\ref{fig1}e):
\begin{equation}
\label{eq.TripleSquareMagnBarrier}
A(x) = \begin{cases} 0, & x < -L \ \text{or}\ x > 2L, \\ -B(x+L), & -L \leq x < 0, \\ 2B(x-L/2), & 0 \leq x < L, \\ B(2L-x), & L \leq x \leq 2L. \end{cases}
\end{equation}

First, we consider a gapless graphene. The structure now has a non-vanishing transparency for all energies, but when $E < E_\mathrm{m}$ with a minimum energy~$E_\mathrm{m}$ given by~\eqref{eq.Emin}, all modes under the right electric barrier become evanescent. The analytical expression for the pumping response is rather complicated, and below we present the numerical results.

\begin{figure}
	\includegraphics[width = .99\columnwidth]{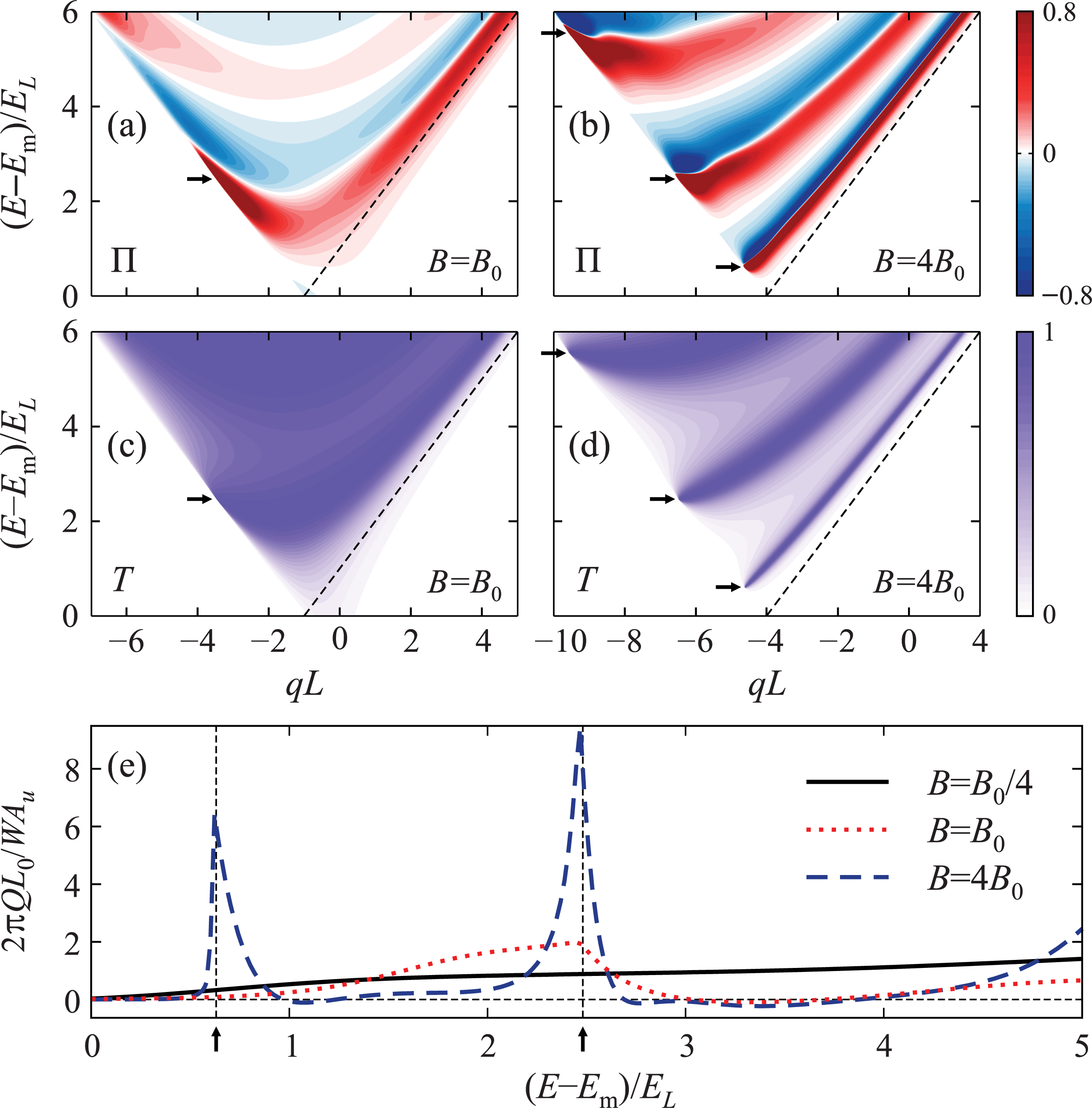}
	\caption{(a)--(d) Contour plot of the pumping response~$\Pi(E,q)$ and the transmission~$T(E,q)$ for the double $\delta$-function magnetic barrier~\eqref{eq.DoubleDeltaMagnBarrier} in graphene with gapless spectrum as a function of the canonical momentum~$q$ and the Fermi energy~$E$. (e)~The pumped charge~$Q(E)$ for a wide ribbon as a function of the Fermi energy~$E$ for different magnetic field strengths.}
	\label{fig6}
\end{figure}

The pumping response, the transmission and the pumped charge are shown in Fig.~\ref{fig6}. In contrast to a single $\delta$-function barrier (cf. Fig.~\ref{fig2} and Fig.~\ref{fig3}), there appear the pronounced resonances in the transmission due to the presence of a additional magnetic barrier at $x=L$~\cite{RVMP08}. The similar behavior is exhibited by the pumping response, and its resonances are accompanied by the resonances of the transmission. The vector potential is now left-right symmetric ($A(x) = A(-x)$) rather than antisymmetric, and the symmetry of the distributions~$T_\tau(q)$ and $\Pi_\tau(q)$ with respect to the interchanging of the sign of the canonical momentum, $q \to -q$, is broken.

As can be seen in Fig.~\ref{fig6}e, the pumped charge~$Q(E)$ is a peaked function of the Fermi energy (peaks become more pronounced as the magnetic field increases). The major contribution to~$Q(E)$ comes from the quasi-bound states between the two $\delta$-function barriers with small longitudinal momentum~$k=k_l=k_r$ in the leads. Hence, the positions of the peaks approximately coincide with the energies~$E_n$ of such states:
\begin{equation}
E_n - E_\mathrm{m} = E_L \left( \frac{\ell_B}{L} \right)^2 \frac{\pi^2n^2}{4}, \quad n=1,2,\ldots,
\end{equation}
which are determined by the simple condition~$k_1L=\pi n$ with $k_1$ the longitudinal momentum between the magnetic barriers. These values are marked with the black arrows in Fig.~\ref{fig6}.

In considering the pumping of the valley current in a gapped graphene, we first note that if the vector potential is left-right symmetric ($A(x) = A(-x)$), the Hamiltonian~\eqref{eq.HamiltonianGap} satisfies the symmetry $\mathcal{O} H(B) \mathcal{O}^{-1} = H(B, U_l \leftrightarrow U_r)$ with $\mathcal{O}=\tau_y \sigma_y R_y$. Then, as it was demonstrated in Ref.~\cite{Zha12}, the transmission is the same for both valleys, $T_\tau(q) = T_{-\tau}(q)$. It can be shown that the pumping response~\eqref{eq.Pi} is also valley-independent, $\Pi_\tau(q) = \Pi_{-\tau}(q)$. This symmetry is realized, e.\,g., for a symmetric double $\delta$-function barrier $B_z = BL\left[\delta(x+L) - \delta(x-L)\right]$ or a double square barrier $B_z = B\left[\sgn(x+L)-2\sgn(x)+\sgn(x-L)\right]$. No valley current will be pumped in this case. If only one of the electrodes has finite magnetization, this symmetry is broken.

\begin{figure}
	\includegraphics[width = .99\columnwidth]{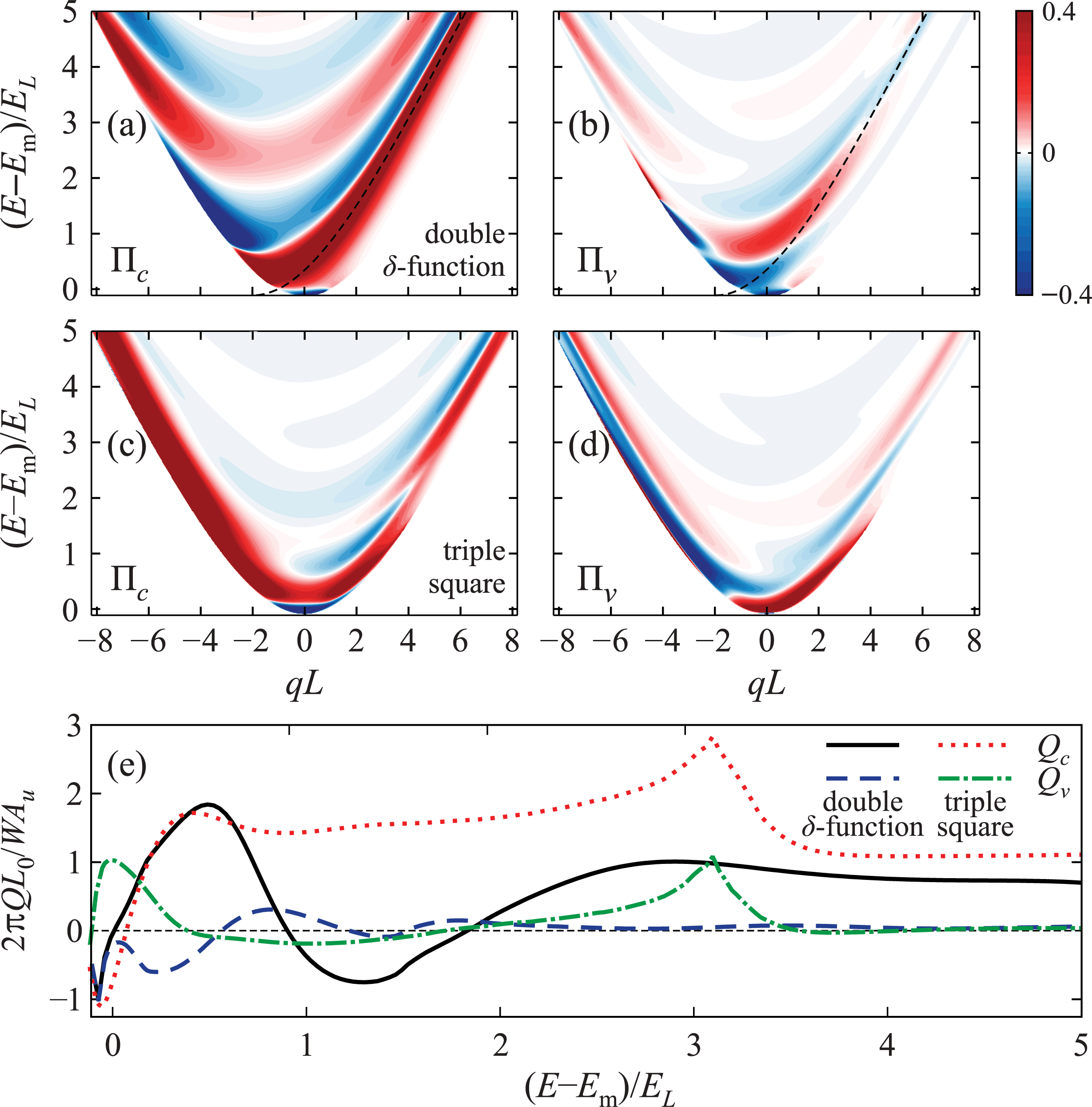}
	\caption{(a)--(d) Contour plot of the total~$\Pi_c(E, q)$ and the valley~$\Pi_v(E,q)$ pumping responses for (a)--(b)~the double $\delta$-function magnetic barrier~\eqref{eq.DoubleDeltaMagnBarrier} and (c)--(d)~the triple square barrier~\eqref{eq.TripleSquareMagnBarrier} in graphene with gapped spectrum ($\Delta=4E_L$, $B=B_0$) as a function of the canonical momentum~$q$ and the Fermi energy~$E$. (e)~The total~$Q_c(E)$ and the valley~$Q_v(E)$ pumped charges for a wide ribbon as a function of the Fermi energy~$E$. Constant offset of electric barriers heights is $U_{l0} = U_{r0} = -0.3E_L$.}
	\label{fig7}
\end{figure}

The results for a gapped graphene with finite offset of electric barriers heights $U_{l0}=U_{r0}=-0.3E_L$ are collected in Fig.~\ref{fig7}. The pumped charges~$Q_c(E)$ and~$Q_v(E)$ exhibit the oscillatory sign-changing behavior, and similar to the structure with a single square barrier considered in the previous section, at some values of the Fermi energy a pure valley current is generated.

The dashed lines in Fig.~\ref{fig6}a--d and Fig.~\ref{fig7}a--b indicate the boundary between propagating and evanescent modes under the right electric barrier (in the region $0<x<L$). We see that the contribution of evanescent modes to pumping is small, but finite. In the case of a gapped graphene only these modes participate in pumping at energies $\Delta < E < E_\mathrm{m}$.

\section{Conclusion}

In this paper we have applied the scattering matrix approach to the adiabatic quantum pumping in a graphene ribbon with a magnetic barrier. By using a $\delta$-function approximation for a magnetic barrier profile, an analytical solution is derived. We find that a finite magnetic field breaks the perfect Klein tunneling so that all propagating modes become sensitive to pumping. At the same time a magnetic barrier decreases the overall efficiency of a quantum pump.

The joint use of a magnetic barrier and band gap engineering in graphene gives a way to generate valley-polarized currents in graphene-based quantum pumps. The parameters of a device can be adjusted such that a pure valley current is produced. The pumping is sensitive to heights~$U_{l0}$ and $U_{r0}$ of electric barriers, and experimentally it may be easier to vary these parameters rather than a band gap value or a magnetic field strength.

A $\delta$-function and a square magnetic barrier profiles, which are employed in this paper and widely used for studying the electron transport through magnetic barriers, can be viewed as simplified approximations to those created experimentally by ferromagnetic strips. More realistic smooth profiles can be analyzed in the same framework. It is interesting to consider pumping outside the bilinear regime where higher valley currents are expected to be obtainable. The motion of an electron in a magnetic field is also affected by its spin, which should be included into the model. These aspects will be analyzed in the future work.

The considered pump effect might be found useful in the field of graphene valleytronics, e.\,g., as a source of valley-polarized and pure valley currents. Experimentally they could be detected using, for example, the valley Hall effect~\cite{XYN07}.

\begin{acknowledgement}
We gratefully acknowledge financial support from the Russian Fund for Basic Research, project No.~10-02-00399 and from the Ministry of Education and Science of the Russian Federation, project No.~8364.
\end{acknowledgement}

\bibliographystyle{epj}
\bibliography{graphene_pump}

\end{document}